\def\npb{Nucl. Phys. B}
\def\plb{Phys. Lett. B}
\def\prl{Phys. Rev. Lett. }

\def\be{\begin{equation}}
\def\ee{\end{equation}}
\def\ba{\begin{eqnarray}}
\def\ea{\end{eqnarray}}

\documentstyle[12pt]{article}
\begin{document}

\title{{\bf Four leptons final states from $\gamma\gamma$ fusion.}}

\author{
{\bf Mauro Moretti,}
\\ \it (e-mail: moretti@vaxfe.fe.infn.it)
}
\date{}
\maketitle

\begin{abstract}
I present a systematic study of all possible four leptons
final states from $\gamma\gamma$ collisions. It is given
a  detailed
account of fermion masses effects 
which are  sizable since several
collinear and $t$ channel enancements occur.
The effects of angular cuts on the final charged leptons
are also discussed.
To perform the computation I have used the
recently developed ALPHA algorithm (and the resulting code)
for the automatic computation of scattering amplitudes.
\end{abstract}



\def \nfigmauro#1{\ifx#1\undefined
\fi}

\def\feyn{1}
\def\xsccu{2}
\def\xsccd{3}
\def\dre{4}
\def\xsncn{5}
\def\fourcharg{6}

\section{Introduction}

It appears now technically 
feasible to operate
the future high energy $e^+ e^-$ colliders \cite{nlc}
 in the $e\gamma$
and $\gamma\gamma$ mode \cite{gcol}.
This last possibility will allow
a detailed study of the non abelian nature of the
electroweak interactions
\cite{selfbos}:
triple and quartic gauge boson couplings as well
as the coupling of gauge bosons with the
higgs particle if it is light enough to be produced.

One of the most important
processes to be studied will be 
W pair production.
At high energies, the cross section is dominated
by $t$ channel virtual W exchange and
becomes nearly
 constant  for a center of mass
energy above $400$ GeV,
with a value
of about $90$ picobarns.
 With the aimed luminosity in the range of
 $10 \div 20$ inverse femtobarns per year
about one millions of W pairs are expected.

Since the W boson decays within the detector
the experimental signature for W pair production
is via its decay products, 
mostly four fermions in the final state.
In view of the forecasted production rate,
to address precision studies
 it is necessary
 to compute the 
rate for the process $\gamma\gamma\rightarrow 4 \ fermions$.

In the paper \cite {gg},
the process
$
\gamma \gamma \rightarrow \bar \nu_e e^-
u \bar d
$
has been studied. 
In this paper I present a systematic study
of leptonic four fermion final states at
a $\gamma\gamma$ collider including the effect of 
fermion masses.

\section{The computation}

The  amplitude for the processes in tables
\ref{xxcc}, \ref{xxnca} and  \ref{xxncb}
is computed using a new technique which,
in collaboration with {\it F. Caravaglios} \cite{alpha},
I have recently developed.
Exploiting the relation between the one-particle irreducible Green 
Functions generator $\Gamma$ and the connected Green Functions
generator $G$ we have proposed a simple numerical algorithm to compute
tree level scattering amplitudes.
We have then implemented the algorithm in a FORTRAN code ALPHA
which presently uses the standard electroweak lagrangian (QCD is
not included yet) and can compute any scattering
amplitude in this framework,
in a fully automatic way.  

The tests we have performed as well as the computations we have done
using the ALPHA code are described in the previous papers
\cite{gg,alpha,lep200}, it is only worth noticing here that, 
because of the automatic approach to the calculation,
there is no need to check the matrix elements for mistakes and bugs.

The input values,
which will be used in
the present paper,
 for particles masses, widths and
for the electroweak coupling constants are reported in table \ref {ew}.
The gauge bosons propagator
$\pi_B$ is taken
as
\be
\pi_B^{\mu\nu} = \frac { - i (g^{\mu\nu} - p^\mu p^\nu 
/M_B^2) }
{p^2 - M_B^2 + i \Gamma_B p^2 \theta (p^2)  /M_B}
\label{rw}
\ee
where $p$ is the gauge boson four momentum, $M_B$ and $\Gamma_B$ are
the gauge bosons mass and width respectively,
and $\theta(p^2)$ is the Riemann $\theta$ function:
it is equal to one for positive $p^2$
and zero otherwise.

\begin{table}
\begin{center}
\begin{tabular}{|c|c|c|c|}
\hline\hline
$M_W=80.23$ GeV & 
$\Gamma_W = 2.03367 $ & 
$M_Z=91.1888$ GeV & 
$\Gamma_Z = 2.4974 $ 
\\ 
\hline 
$ \sin^2 \theta_W=0.23103$ &
$\alpha_{QED}=1./128.07$ & 
$m_e=0.51$  MeV & 
$m_\mu=105.66$  MeV 
\\ 
\hline 
$m_\tau=1.7771$  GeV & 
  & &   \\
\hline  \hline
\end{tabular}
\end{center}
\caption {
Input parameters for the electroweak lagrangian.
$m_e$,  $m_\mu$, $m_\tau$  are the electron muon and tau 
masses respectively, $M_W$, $M_Z$,
 $\Gamma_W$ and  $\Gamma_Z$
the W and Z bosons
masses and widths, $\theta_W$ is the Weinberg angle
and $\alpha_{QED}$ is the electromagnetic coupling constant.
Tree level relationships among the parameters of
the standard model electroweak lagrangian are assumed.
}
\label{ew}
\end{table}

\nfigmauro\feyn
Although the ALPHA algorithm does not make use
of the Feynman  graphs technique to compute
the scattering  matrix elements,
it is useful to refer to the Feynman graphs
to discuss the main kinematical and dynamical features
for the processes under study. The relevant 
Feynman graphs are given in fig.~\feyn.

The two 
\footnote {
Actually, in the case of interest here, the diagram $ra$ 
represents $two$ diagrams rather than $one$,
since, exchanging the two external photon lines,
one obtains two inequivalent diagrams.
This is essential in doing the computation
but does not affect the present discussion.
Therefore, in  the following, I will not pay
attention to the real number 
of Feynman diagrams but rather to the relevant topologies.
}
diagrams $ra$ and $rb$
probe the non abelian nature of the Electroweak interactions
and involve both triple and quartic gauge bosons self couplings.
They describe the production and decay of a pair of on/off shell W.
From the form of the propagator of the W boson (\ref{rw})
it is clear that most of the contribution,
of the diagrams $ra$ and $rb$  to the cross section,
occurs for almost on shell W.
Because of the $t$ channel virtual W exchange in the diagram $ra$,
for relativistic W, the cross section is strongly peaked for 
W pairs emitted along the beam direction.

The contribution of the diagram $c$ 
is more important when the internal gauge boson (W, Z or $\gamma$)
is almost on shell and, for light fermions,
when one of the external charged fermions is
emitted  collinear to the beam direction.
In this case, in fact, one of the internal fermions is nearly on shell
and this leads to a logarithmic enancement
of the
cross section.

The diagram $ct$ has the same `singularities' 
\footnote {
Actually the cross sections for the processes I am discussing
are finite at the tree level, and the word `singularity'
is always used in a loose sense, to denote a small region
in the phase space which gives an important contribution
to the cross section.
}
which have been
discussed for the $c$ diagram and has an additional
$t$ channel `singularity' when the decaying $W$ boson is emitted
collinear to the beam.

The diagram $cc$ exhibits a double collinear `singularity'
as well as a $t$ channel virtual gauge boson exchange
with the consequent enancement.

Finally the $cf$ diagram has both a doubly collinear `singularity'
as well as the enancement due to the production of an almost on shell 
gauge boson.

Due to the large 
variety of peaking behaviors of the matrix element,
to perform the numerical integration over
the phase space variables, one
 needs to increase  the sampling in the relevant phase space regions.
To this purpose  I have used the
package VEGAS \cite{vega}
and all the reported results are obtained with at least
twenty VEGAS estimates of the integral with a $\chi^2$
smaller than two. To properly describe the most important
phase space regions it has been necessary to split
the integration domain into three different regions
and to use a suitable set of phase space 
variables in each of these regions.

The performances of the ALPHA code as 
an event generator for $\gamma\gamma$ processes 
are discussed in \cite{gg}
and will not be repeated here.

\section{The results}
I will discuss separately
the processes which do involve virtual $W$
exchange and which I will call CC ({\it Charged Current})
processes in the following and those which proceed only via
neutral gauge bosons exchange which will be referred as NC 
({\it Charged  Current}) processes
in the following.
\subsection{CC processes}
This class of processes can be divided
in two additional subclasses:
the processes $CC_{ee}$, $CC_{\mu\mu}$
and $CC_{\tau\tau}$ (see table \ref{xxcc}
for the definition of $CC_{jk}$)
where all the final leptons 
belongs to the same family 
and the processes
$CC_{e\mu}$, $CC_{e\tau}$
and $CC_{\mu\tau}$.
The diagrams
$ra$, $rb$, $c$, $ct$ and  $cc$ 
do contribute to all $CC$ processes
whereas the diagram $cf$ contributes
only to single leptonic flavour final states:
in this case the virtual gauge boson is a Z boson decaying into 
a neutrino pair.

\nfigmauro\xsccu
The cross section as a function of various angular cuts
is plotted in fig.~\xsccu \ 
and a few values for some angular cuts are reported
in table \ref{xxcc}.

\begin{table}
\begin{center}
\begin{tabular}{|c|c|c|c|c|c|c|}
\hline
Final State & Label & Energy & $\sigma(0.9)$ & $\sigma(0.98) $ &
$\sigma(0.9975) $& $\sigma(1) $\\ 
\hline\hline
$e^+e^-\bar\nu_e\nu_e $& $CC_{ee}$ & 300 &
0.6537(8) & 0.9096(9) & 1.001(1) & 1.200(2)
\\ \hline
$\mu^+\mu^-\bar\nu_\mu\nu_\mu $& $CC_{\mu\mu}$ & 300 &
0.6533(6) & 0.9095(7) & 1.0011(8) & 1.0767(9)
\\ \hline
$\tau^+\tau^-\bar\nu_\tau\nu_\tau$& $CC_{\tau\tau}$ & 300 &
0.6508(5) & 0.9045(5) & 0.9921(6) & 1.0201(8)
\\ \hline
$e^+e^-\bar\nu_e\nu_e$ & $CC_{ee}$ & 500 &
0.456(1) & 0.876(1) & 1.091(2) & 1.415(2)
\\ \hline
$\mu^+\mu^-\bar\nu_\mu\nu_\mu$& $CC_{\mu\mu}$ & 500 &
0.4555(7) & 0.8761(9) & 1.090(1) & 1.227(1)
\\ \hline
$\tau^+\tau^-\bar\nu_\tau\nu_\tau$& $CC_{\tau\tau}$ & 500 &
0.4536(7) & 0.8718(9) & 1.083(1) & 1.150(1)
\\ \hline
$e^+e^-\bar\nu_e\nu_e$& $CC_{ee}$ & 1000 &
0.1490(4) & 0.545(1) & 1.024(2) & 1.502(5)
\\ \hline
$\mu^+\mu^-\bar\nu_\mu\nu_\mu$& $CC_{\mu\mu}$ & 1000 &
0.1491(4) & 0.5466(9) & 1.026(1) & 1.313(2)
\\ \hline
$\tau^+\tau^-\bar\nu_\tau\nu_\tau$& $CC_{\tau\tau}$ & 1000 &
0.1486(4) & 0.543(1) & 1.012(1) & 1.191(2)
\\ \hline
$e^+\mu^-\bar\nu_\mu\nu_e$& $CC_{e\mu}$ & 300 &
0.6520(7) & 0.9065(8) & 0.9958(8) & 1.093(1)
\\ \hline
$e^+\tau^-\bar\nu_\tau\nu_e$& $CC_{e\tau}$ & 300 &
0.6509(5) & 0.9047(6) & 0.9923(6) & 1.0689(7)
\\ \hline
$\mu^+\tau^-\bar\nu_\tau\nu_\mu$& $CC_{\mu\tau}$ & 300 &
0.6511(6) & 0.9054(7) & 0.9928(7) & 1.0332 (2)
\\ \hline
$e^+\mu^-\bar\nu_\mu\nu_e$& $CC_{e\mu}$ & 500 &
0.4555(6) & 0.8754(9) & 1.089(1) & 1.276(1)
\\ \hline
$e^+\tau^-\bar\nu_\tau\nu_e$& $CC_{e\tau}$ & 500 &
0.4544(5) & 0.8733(7) & 1.085(1) & 1.239(2)
\\ \hline
$\mu^+\tau^-\bar\nu_\tau\nu_\mu$& $CC_{\mu\tau}$ & 500 &
0.4545(4) & 0.8733(6) & 1.0840(7) & 1.1797(9)
\\ \hline
$e^+\mu^-\bar\nu_\mu\nu_e$& $CC_{e\mu}$ & 1000 &
0.1496(4) & 0.548(1) & 1.024(1) & 1.405(2)
\\ \hline
$e^+\tau^-\bar\nu_\tau\nu_e$& $CC_{e\tau}$ & 1000 &
0.1489(3) & 0.5450(5) & 1.0211(8) & 1.339(2)
\\ \hline
$\mu^+\tau^-\bar\nu_\tau\nu_\mu$& $CC_{\mu\tau}$ & 1000 &
0.1491(3)  & 0.5448(6) & 1.0219(8) & 1.262(3)
\\ \hline
$Resonant$ & Approximation & 300 &
0.6610(3) & 0.9073(4) & 0.9823(4) & 0.9940(4)
\\ \hline
$Resonant$ & Approximation & 500 &
0.4770(6) & 0.8962(8) & 1.0893(8) & 1.1244(9)
\\ \hline
$Resonant$ & Approximation & 1000 &
0.1671(1) & 0.5847(3)  & 1.0568(4) & 1.1923(4) 
\\ \hline
\hline
\end{tabular}
\end{center}
\caption {
Cross sections
for $CC$ processes for center of mass energies of 300, 500 and
1000 GeV. $\sigma (x)$ is the cross section
when the angular cut $ | \cos \theta_f | < x $
is applied
and 
$\theta_f$ 
is defined as follows:
$ | \cos \theta_f | = {\mathrm Min} \{ | \cos \theta_{l^+} |,
| \cos \theta_{l^-} | \} $
where $\theta_{l^-}$ and $\theta_{l^+} $
are the
the angles of the negatively and positively
charged leptons respectively, with the beam direction.
Energies are in GeV and cross sections in picobarns.
}
\label{xxcc}
\end{table}

The $CC_{ee}$, $CC_{\mu\mu}$
and $CC_{\tau\tau}$ processes
differ among each other only for the mass
of the final charged leptons.
The effect of fermion masses is
manifest.
The value of the total (without any cuts)
cross section is sensitive to fermion masses,
in fact the cross section for the
$CC_{\mu\mu}$
and $CC_{\tau\tau}$
processes is lower than that of $CC_{ee}$
process by a factor of about 12\% and 17\% 
respectively.
The difference is mostly due to the contribution
of the diagrams $c$, $cc$ and $cf$.
All these diagrams are divergent, in the limit
of massless fermions,
when a charged leptons is emitted collinear to the beam.
Lepton masses act as a physical cut-off for this collinear logarithmic
`divergence'
and all the cross sections
receive a contribution proportional to 
$\log (m_l/E_\gamma)$
($m_l$ is the mass of the relevant lepton).
As it can be seen from table \ref{xxcc},
if an  angular cut is imposed, forcing 
a small acollinearity with respect to the beam direction
for the 
charged fermions,
fermion masses become almost unimportant since
the angular cut 
regulates
the `divergence' more efficiently
than the lepton masses
and the leading contribution
of collinearly `divergent diagrams'
becomes, irrespectively of the fermion masses,
 proportional
to $\log (1-\cos\theta_c)$,
$\theta_c$ being the imposed angular cut.

Even after angular cuts are imposed,
a small difference remains among $CC_{\tau\tau}$
and $CC_{\mu\mu}$, $CC_{ee}$ processes.
Althoug, as it can be seen from fig.~\xsccu,
the accuracy of the computation is not entirely adequate
to discuss such a small effect, it appears
that the difference is of the order of a few per
mille.
From purely
kinematical considerations one expects an effect of order one per
mille: in fact,
if one computes the volume of
the phase space for the decay of
a W boson into a pair of fermions,
there is a correction of $m_l^2/(2 M_W^2)$
($m_l$ being the lepton mass)
to the result which is obtained assuming  massless fermions
and, for the $\tau$ lepton,
this correction is indeed of order one per mille.
In table \ref{difetau} I report a few values of the 
difference among the cross
sections of $CC_{ee}$ and $CC_{\tau\tau}$ processes
as a function
of several angular and energy cuts. Although,
as already noticed, the accuracy of the calculation
does not allow a definite conclusion, 
it seems that the effect disappears when
charged fermions energies are greater than $5 \div 10 $ GeV
and it is slightly weaker when the fermions are emitted at larger angles
with respect to the beam.
Therefore it seems that, unless a cut of order
$10$ GeV is imposed on the energies of final leptons,
one needs to account also for the dynamical effect
of the $\tau$ mass.

\begin{table}
\begin{center}
\begin{tabular}{|c|c|c|c|c|c|c|}
\hline
Energy Cut & 
$\Delta\sigma_{e\tau}(0.9975) $ & 
$\Delta\sigma_{e\tau}(0.995) $ & 
$\Delta\sigma_{e\tau}(0.99) $ & 
$\Delta\sigma_{e\tau}(0.98) $ & 
$\Delta\sigma_{e\tau}(0.92) $ &
$\sigma_{ee}(0.9975) $ 
\\ 
\hline\hline
$E_l > 0 $ GeV & 7.0(1.2) & 5.8(1.2) & 4.3(1.2) & 3.4(1.1) &1.2(0.9)
& 1091(1)
\\ \hline
$E_l > 3 $ GeV & 5.3(1.2) & 4.3(1.2) & 3.0(1.2) & 2.4(1.1) &0.73(0.93)
& 1089(1)
\\ \hline
$E_l > 9 $ GeV & 2.4(1.2) & 1.8(1.2) & 0.9(1.2) &0.7(1.1) & -0.09(0.9)
& 1054(1)
\\ \hline
\hline
\end{tabular}
\end{center}
\caption {
Difference, in femtobarns, among the cross sections for 
$CC_{ee}$ and $CC_{\tau\tau}$ processes
for a center of mass energy of 500 GeV.
$E_l$ is the lowest energy of final charged leptons, 
$\Delta\sigma_{e\tau} (x)$ is the difference
when the angular cut $ | \cos \theta_f | < x $
is applied
and 
$\theta_f$ 
is defined as follows:
$ | \cos \theta_f | = {\mathrm Min} \{| \cos \theta_{l^+} |,
| \cos \theta_{l^-} |\} $
where $\theta_{l^-}$ and $\theta_{l^+} $
are the
the angles of the negatively and positively
charged leptons respectively, with the beam direction.
$\sigma_{ee}$ is the cross section for the $CC_{ee}$ process.}
\label{difetau}
\end{table}

The $CC_{e\mu}$, $CC_{e\tau}$
and $CC_{\mu\tau}$ processes
differ among each other only for the mass
of the final charged leptons.
They do not receive contribution from the $cf$
diagram and the $c$ diagram is possible only when the 
internal gauge boson is a W.
\nfigmauro\xsccd
The cross section as a function of the angular cut
is plotted in fig.~\xsccd \ and some numerical values
are reported in tab \ref{xxcc}.
It is manifest that, with the exception of the already noticed
effect of fermion masses in the case of nearly collinear fermion
emission, there is no difference, at the per mille level,
with single flavour final states.
This demonstrates that, when a small angular cut is applied,
the Z exchange contribution (diagrams $cf$ and $c$) is
very small as it can also be seen looking at the rate for the $NC_{e}$
process (see table \ref{xxnca}).

\nfigmauro\dre
The main contribution 
to the cross section
of $CC$ processes 
comes from W pair production and decay
(diagrams $ra$ and $rb$ in fig.~\feyn).
As already discussed in \cite{gg} both the narrow width approximation
and
the approximation (which will be referred as $resonant$ approximation
in the following)
based on the subset of 
doubly resonant diagrams $ra$ and $rb$
are inadequate for precision studies.
In fig.~\dre \ I plot the relative difference
among the complete calculation for the $CC_{ee}$ process
and the $resonant$ approximation:
the difference is sizable 
with any angular cut with the exception
of an accidental cancellation at a specific (and
energy dependent) value of the angular cut.
A few values of the cross section, calculated in the {\it resonant}
approximation, are reported in table \ref{xxcc}.
The discrepancy increases with the beam energy and 
this corroborates the hypothesis that
the effect is due to the lack of gauge invariance
of the $resonant$ approximation
and to the related unitarity violation.

\subsection{NC processes}
Let us first discuss those processes involving a final
neutrino pair $NC_e$, $NC_\mu$ and $NC_\tau$.
They all proceed via virtual, on/off shell, Z boson exchange
(diagrams $c$ and $cf$).
\nfigmauro\xsncn
The cross section is plotted in fig.~\xsncn
\ as a function of the angular cut
and a few values are reported in table \ref{xxnca}. 
There is again a significative 
effect of fermion masses for fermion emission at very 
small angles with respect to the beam direction
and the difference disappears for acollinear fermions.
The size of the cross sections for these processes
provides an indirect confirmation of the observation
made for $CC$ processes for which
Z exchange contribution appears to be 
negligible at the level of accuracy of the present computation.

\begin{table}
\begin{center}
\begin{tabular}{|c|c|c|c|c|c|c|}
\hline
Final State & Label & Energy & $\sigma(0.9)$ & $\sigma(0.98) $ &
$\sigma(0.998) $& $\sigma(1) $\\ 
\hline\hline
$e^+e^-\bar\nu_\tau\nu_\tau $& $NC_{e}$ & 300 &
0.93(1) & 2.08(2)  & 4.57(3)  & 
62.93(8) \\ \hline
$\mu^+\mu^-\bar\nu_\tau\nu_\tau $& $NC_{\mu}$ & 300 &
0.943(5) & 2.105(8) & 4.57(1) & 20.89(2)
\\ \hline
$\tau^+\tau^-\bar\nu_\mu\nu_\mu $& $NC_{\tau}$ & 300 &
0.900(5)& 1.979(7) & 4.07(1) & 7.83(1)
\\ \hline
$e^+e^-\bar\nu_\tau\nu_\tau $& $NC_{e}$ & 500 &
0.548(6) & 1.9275(9) &2.570(1)  & 
41.41(4) \\ \hline
$\mu^+\mu^-\bar\nu_\tau\nu_\tau $& $NC_{\mu}$ & 500 &
0.546(4) & 1.191(7) & 2.58(1) & 14.54(2)
\\ \hline
$\tau^+\tau^-\bar\nu_\mu\nu_\mu $& $NC_{\tau}$ & 500 &
0.537(3) & 1.164(3) & 2.441(5) & 5.95(1)
\\ \hline
$e^+e^-\bar\nu_\tau\nu_\tau $& $NC_{e}$ & 1000 &
0.249(2)& 0.508(4) & 1.025(6)  & 
18.78(2) \\ \hline
$\mu^+\mu^-\bar\nu_\tau\nu_\tau $& $NC_{\mu}$ & 1000 &
0.250(1) & 0.512(2) & 1.034(3) & 7.003(6)
\\ \hline
$\tau^+\tau^-\bar\nu_\mu\nu_\mu $& $NC_{\tau}$ & 1000 &
0.246(1) & 0.507(2) & 1.018(2) & 3.144(4)
\\ \hline
\hline
\end{tabular}
\end{center}
\caption {
Cross sections
for $NC$ 
processes for center of mass energies of 300, 500 and
1000 GeV. Only final states with a neutrino pair.
$\sigma (x)$ is the cross section
when the angular cut $ | \cos \theta_f | < x $
is applied
and 
$\theta_f$ 
is defined as follows:
$ | \cos \theta_f | = {\mathrm Min} \{| \cos \theta_{l^+} |,
| \cos \theta_{l^-} |\} $
where $\theta_{l^-}$ and $\theta_{l^+} $
are the
the angles of the negatively and positively
charged leptons respectively with the beam direction.
Energies are in GeV and cross sections in femtobarns.
}
\label{xxnca}
\end{table}

The production of four charged leptons proceeds
via the 
exchange of virtual on/off shell Z bosons and photons (diagrams $c$,
$cf$ and $cc$).
When no cut is applied on the final leptons 
 the cross
section is dominated by the $t$ channel virtual photon exchange
of diagram $cc$. 
This diagram is singular if the virtual photon
become massless and the only cut-off to this `singularity'
is provided by the fermion masses. 
The expected behavior of the cross section is therefore
$\sigma \sim 1/m_f^2$ times the logarithmic enancment
due to the collinear `singularities'
associated with internal fermions lines.
Especially for $e^+e^-e^+e^-$
final state the cross section is huge.
\begin{table}
\begin{center}
\begin{tabular}{|c|c|c|c|c|c|c|}
\hline
Energy Cut  & $\sigma(\alpha_3,\beta_2)$ &
$\sigma(\alpha_3,\beta_1)$ &
$\sigma(\alpha_2,\beta_2)$ &
$\sigma(\alpha_2,\beta_1)$ &
$\sigma(\alpha_1,\beta_2)$ &
$\sigma(\alpha_1,\beta_1)$ 
\\
\hline\hline
$E_l>0$ GeV & 4.08(11) & 4.70(12) & 8.29(22) & 9.74 (23) & 136(1) & 661(2)
\\ \hline
$E_l>2$ GeV & 4.07(11) & 4.70(12) & 8.28(22) & 9.74 (22) & 136(1) & 660(2)
\\ \hline
$E_l>4$ GeV & 3.99(11) & 4.60(12) & 8.13(22) & 9.57 (22) & 132(1) & 649(2)
\\ \hline
$E_l>6$ GeV & 3.84(10) & 4.45(12) & 7.90(22) & 9.32 (22) & 127(1) & 633(2)
\\ \hline
$E_l>9$ GeV & 3.68 (10) & 4.27(12) & 7.59(21) & 8.97(22) & 119(1) & 605(2)
\\ \hline
\hline
\end{tabular}
\end{center}
\caption {
Cross sections
for $NC_{\tau\tau}$ process for a center of mass energies of 500 GeV.  
$\sigma (x_1,x_2)$ is the cross section
when the angular cuts $ | \cos \theta_f | < x_1 $
$ | \cos \theta_p | < x_2 $    
are applied
and 
$\theta_f$, $\theta_p$ 
are  defined as follows:
$ | \cos \theta_f | = {\mathrm Min} \{| \cos \theta_{j} |
\} $
and $ | \cos \theta_p | = {\mathrm Min} \{\cos\theta_{j,k} \}$
where $\theta_{j}$ 
are the
the angles of the 
charged leptons with the beam direction
and $\theta_{j,k}$ are the angles among each final lepton pairs.
The angles $\alpha_1$, $\alpha_2$, $\alpha_3$, $\beta_1$ and
$\beta_2$ are equal to $0^o$, $5.89^o$, $8.61^o$, $0^o$ and $8.33^o$
degrees respectively and $E_l$ is the lowest among final leptons
energies.
Cross sections are in femtobarns.
}
\label{xxnctau}
\end{table}
In table \ref{xxnctau}
I give a few values of the cross section for the $\tau^+\tau^-\tau^+\tau^- $
final state as
a function of various angular cuts. As expected
most of the 
cross section 
occurs for final leptons collinear to the beam and very
close (in direction) to each other.

\begin{table}
\begin{center}
\begin{tabular}{|c|c|c|c|c|c|c|}
\hline
Final State & Label & Energy & $\sigma(\alpha_2,\beta_2)$ &
$\sigma(\alpha_2,\beta_1)$ &
$\sigma(\alpha_1,\beta_2)$ &
$\sigma(\alpha_1,\beta_1)$ 
\\
\hline\hline
$e^+e^-e^+e^- $& $NC_{ee}$ & 300 &
15.89(46) & 34.2(5.7) & 275(3) & 1174(8)
\\ \hline
$\mu^+\mu^-\mu^+\mu^- $& $
NC_{\mu\mu}$ & 300 &
13.25(23) & 18.93(80) & 266(3) & 1078(6)
\\ \hline
$\tau^+\tau^-\tau^+\tau^- $& $NC_{\tau\tau}$ & 300 &
9.89(42) & 11.04(43) & 145(1) & 320(2) 
\\ \hline
$e^+e^-\mu^+\mu^- $& $NC_{e\mu}$ & 300 &
30.8(6) & 48.1(1.0) & 535(4) & 2292(23) 
\\ \hline
$e^+e^-\tau^+\tau^- $& $NC_{e\tau}$ & 300 &
28.7(1.0) &42.0(1.3)  & 402(4) & 1215(5)
\\ \hline
$\mu^+\mu^-\tau^+\tau^- $& $NC_{\mu\tau}$ & 300 &
24.2(8) & 29.1(8) & 384(2) & 1145(3)
\\ \hline
$e^+e^-e^+e^- $& $NC_{ee}$ & 500 &
6.13(33) & 9.68(71) &   99.46(88)
& 412(2)
\\ \hline
$\mu^+\mu^-\mu^+\mu^- $& $NC_{\mu\mu}$ & 500 &
5.69(21) & 8.28(82) &   97.7(7) & 411(2)
\\ \hline
$\tau^+\tau^-\tau^+\tau^- $& $NC_{\tau\tau}$ & 500 &
4.16(12) & 4.69(12) & 68.39(93) & 188(1)
\\ \hline
$e^+e^-\mu^+\mu^- $& $NC_{e\mu}$ & 500 &
12.28(40) & 18.9(1.4) & 200(2) & 834.0(3)
\\ \hline
$e^+e^-\tau^+\tau^- $& $NC_{e\tau}$ & 500 &
11.2(5) & 16.2(8) & 168(3) & 576.2(3)
\\ \hline
$\mu^+\mu^-\tau^+\tau^- $& $NC_{\mu\tau}$ & 500 &
9.04(21) & 10.7(2) & 166(4) & 555(5)
\\ \hline
$e^+e^-e^+e^- $& $NC_{ee}$ & 1000 &
1.57(7) & 2.55(37) & 27.33(94) & 110(1)
\\ \hline
$\mu^+\mu^-\mu^+\mu^- $& $NC_{\mu\mu}$ & 1000 &
1.56(11) & 2.04(11) & 25.22(29) & 102.6(6)
\\ \hline
$\tau^+\tau^-\tau^+\tau^- $& $NC_{\tau\tau}$ & 1000 &
1.42(11) & 1.62(11) & 20.87(18) & 69.8(2)
\\ \hline
$e^+e^-\mu^+\mu^- $& $NC_{e\mu}$ & 1000 &
3.36(16) & 4.76(18) & 50.70(78) & 215.6(3.6)
\\ \hline
$e^+e^-\tau^+\tau^- $& $NC_{e\tau}$ & 1000 &
3.20(13) & 4.57(25) & 48.8(1.6) & 178.0(1.7)
\\ \hline
$\mu^+\mu^-\tau^+\tau^- $& $NC_{\mu\tau}$ & 1000 &
3.04(15) & 3.81(17) & 47.0(8) & 173.0(1.0)
\\ \hline
\hline
\end{tabular}
\end{center}
\caption {
Cross sections
for $NC$ processes for center of mass energies of 300, 500 and
1000 GeV.  Only final states with four charged leptons in the final state.
$\sigma (x_1,x_2)$ is the cross section
when the angular cuts $ | \cos \theta_f | < x_1 $
$ | \cos \theta_p | < x_2 $    
are applied
and 
$\theta_f$, $\theta_p$ 
are  defined as follows:
$ | \cos \theta_f | = {\mathrm Min} \{| 
 \cos \theta_{j}| \} $
and $  \cos \theta_p = {\mathrm Min} \{\cos\theta_{j,k}  \}$
where $\theta_{j}$ are the
the angles of the 
charged leptons with the beam direction
and $\theta_{j,k}$ are the angles among each final lepton pairs.
The angles $\alpha_1$, $\alpha_2$, $\beta_1$ and
$\beta_2$ are equal to $2^o$, $8.61^o$, $2^o$ and $8.33^o$
degrees respectively.
Energies are in GeV and cross sections in femtobarns.
}
\label{xxncb}
\end{table}

Since the detection unavoidably will discard
such events in table \ref{xxncb} I report
the cross sections for this class of processes
imposing the requirement
that final leptons are emitted with an angle
of at least $2^o$ with respect to the beam
direction and among each other. Althoug this
requirements appears extremely mild 
the effect is manifest: the cross section
is drastically reduced since now the cut-off
on the internal photons virtuality is much harder.
\nfigmauro\fourcharg 
In fig.~\fourcharg \ I plot the cross section for the $NC_{ee}$
process as a function of various angular cuts.

\begin{table}
\begin{center}
\begin{tabular}{|c|c|c|c|c|c|}
\hline
Energy &  Energy Cut & $\sigma(\alpha_2,\beta_2)$ &
$\sigma(\alpha_2,\beta_1)$ &
$\sigma(\alpha_1,\beta_2)$ &
$\sigma(\alpha_1,\beta_1)$ 
\\
\hline\hline
 300 & $E_l > 0 $ GeV &
30.83(60)& 48.1(1.0) & 535(4) & 2292(23)
\\ \hline
 300 & $E_l > 3 $ GeV &
21.92(44) & 30.47(68) & 463(4) & 2028(14)
\\ \hline
 300 & $E_l > 9 $ GeV &
17.09(43) & 22.58(65) & 361(4) & 1619(14)
\\ \hline
 500 & $E_l > 0 $ GeV &
 12.25(40) & 18.9(1.4) & 200(2) & 834(3)
\\ \hline
 500 & $E_l > 3 $ GeV &
 9.56(28) & 14.9(1.4) & 182(2) & 778(3)
\\ \hline
 500 & $E_l > 9 $ GeV &
 8.06(27) & 10.7(3) & 156(2) & 675(3)
\\ \hline
 1000 & $E_l > 0 $ GeV &
 3.36(16) & 4.76(18) &50.70(78) & 216(4)
\\ \hline
 1000 & $E_l > 3 $ GeV &
 3.00(16) & 4.08(17) & 47.48(63) & 207(4)
\\ \hline
 1000 & $E_l > 9 $ GeV &
 2.62(15) & 3.44(15) & 43.71(62) & 193(4)
\\ \hline
\hline
\end{tabular}
\end{center}
\caption {
Cross sections
for $NC_{e\mu}$ process for center of mass energies of 300, 500 and
1000 GeV.  $\sigma (x_1,x_2)$ is the cross section
when the angular cuts $ | \cos \theta_f | < x_1 $
$ | \cos \theta_p | < x_2 $    
are applied
and 
$\theta_f$, $\theta_p$ 
are  defined as follows:
$ | \cos \theta_f | = {\mathrm Min} \{ | \cos \theta_{j} |
\} $
and $ | \cos \theta_p | = {\mathrm Min} \{\cos\theta_{j,k}  \}$
where $\theta_{j}$ 
are the
the angles of the 
charged leptons with the beam direction
and $\theta_{j,k}$ are the angles among each final lepton pairs.
The angles $\alpha_1$, $\alpha_2$, $\beta_1$ and
$\beta_2$ are equal to $2^o$, $8.61^o$, $2^o$ and $8.33^o$
degrees respectively, $E_l$ is the lowest  among final leptons energies.
Energies are in GeV and cross sections in femtobarns.
}
\label{xsnclepen} 
\end{table}

Lepton energies are also relevant and there is a
sizable production of low energy leptons.
In fact although disfavored because of the small phase space
these events are enanced because
lepton pairs with  
low invariant mass and momentum  are produced.
In table \ref{xsnclepen} the cross section as a function
of several angular and energy cuts is
reported for the process $NC_{e\mu}$.

\section{Conclusions}
I have 
used the
recently developed
ALPHA algorithm (and the resulting code)
to perform a systematic study of the
processes $\gamma\gamma\rightarrow 4 \ leptons$
which will be relevant at future $e^+e^-$
colliders when operating in the $\gamma\gamma$
mode. 
\vskip 10pt

For those process which proceed
via virtual W exchange (labelled as $CC_{jk}$ in the text and in
table \ref{xxcc}) the bulk of the cross section
is due to the production and decay of a W pair. Another
important contribution,
at the level of several per cent, arises  because of the 
emission of a collinear fermion and a single W.
To perform precision studies the full computation is therefore
needed.

When no angular cut is imposed fermion masses act as cut-off
of the singularities which occur in correspondence
of charged fermions emitted collinear to the beam direction
and are therefore relevant. If a moderate  angular cut is imposed
fermion masses become almost unimportant. 

Because of 
the contribution of $t$ channel virtual
W exchange of the diagram $ra$ the total rate is nearly constant 
for a center of mass energy above 400 GeV.
When angular cuts are imposed the usual fall-off 
with the center of mass energy is observed.

\vskip 10pt
The reactions which proceed only via neutral gauge bosons
exchange (labelled $NC_{jk}$ in the text and in tables
\ref{xxnca} and \ref{xxncb}) present different features
according to the charge multiplicity of the final state.
\vskip 5pt

If in the final state there are a neutrino and
a charged lepton pairs
the bulk of the cross section is due to the $c$
diagram 
when both of the charged fermions are collinear to the beam
a Z boson is emitted and then decay in a neutrino pair.
The cross section is always below a few femtobarns with the
exception of fermions emitted very close to the beam
direction where fermion mass effects are important and
the cross sections range from a few to
one hundred femtobarns 
\vskip 5pt

If in the final state there are two pairs of charged fermions
the bulk of the contribution comes from the diagram $cc$,
namely the two photon radiates a pair of almost on shell
fermions and two of these  fermions undergo
compton scattering with its characteristic
singularity in the forward direction. Since 
the fermions which scatter have a small virtuality
the typical rate of these processes is proportional
to $1/m_l^2$ times logarithmic enancements.
Because of the smallness of lepton masses the
resulting cross sections are huge and strongly dependent
on angular cuts: a relatively mild angular cut is enough
to reduce the rate by order of magnitudes.

\vskip 10pt
A final comment is in order here:
because of the finite (and {\it running} widht of massive gauge bosons in
(\ref{rw}) the results I have presented are not gauge invariant.
A discussion (incomplete) about this issue can be found in 
\cite{gg} where it is argued that it is likely that this fact
is numerically irrelevant (in the Unitary gauge which is used here).
This fact has anyway to be confirmed by an explicit
computation which must respect gauge invariance as well
as account for gauge bosons widths in a satisfactory way.

\vskip 40pt
\noindent {\Large \bf   Acknowledgments}\\
\vskip 9pt
I thank the {\it Associazione per lo Sviluppo
della Fisica Subatomica,} {\it Ferrara} for financial support
and {\it INFN,} {\it sezione di Ferrara} for making available
computing facilities.


\newpage
\noindent\hskip -7pt
{\Large {\bf Figure Captions}}

\vskip 20pt
\begin{list}{P}{\labelwidth=100pt} 

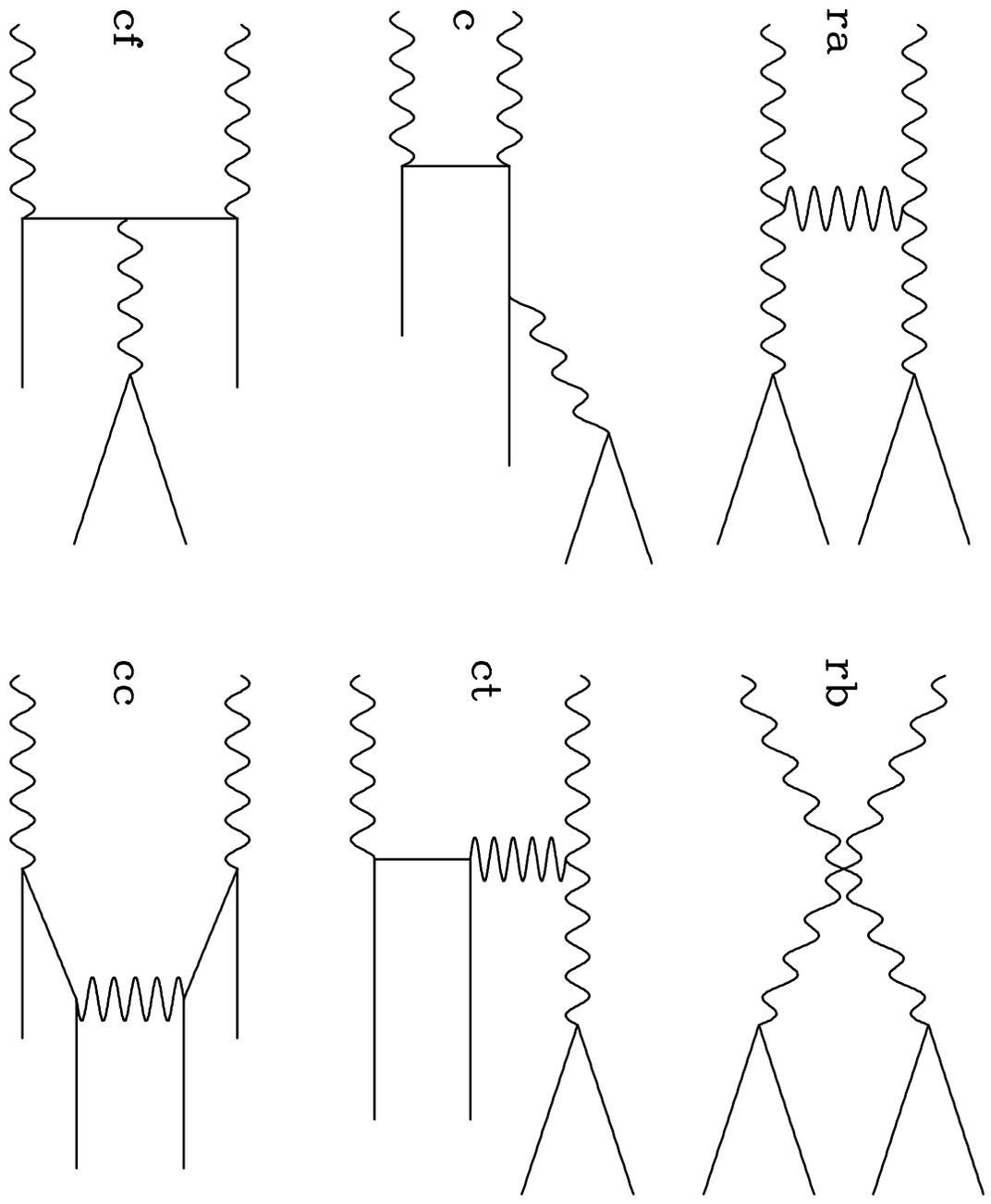
\item [Fig.~1] Feynman diagrams for $\gamma\gamma\rightarrow 4 \ fermions$.
Straight lines represents 
 fermions and wiggled lines gauge bosons.

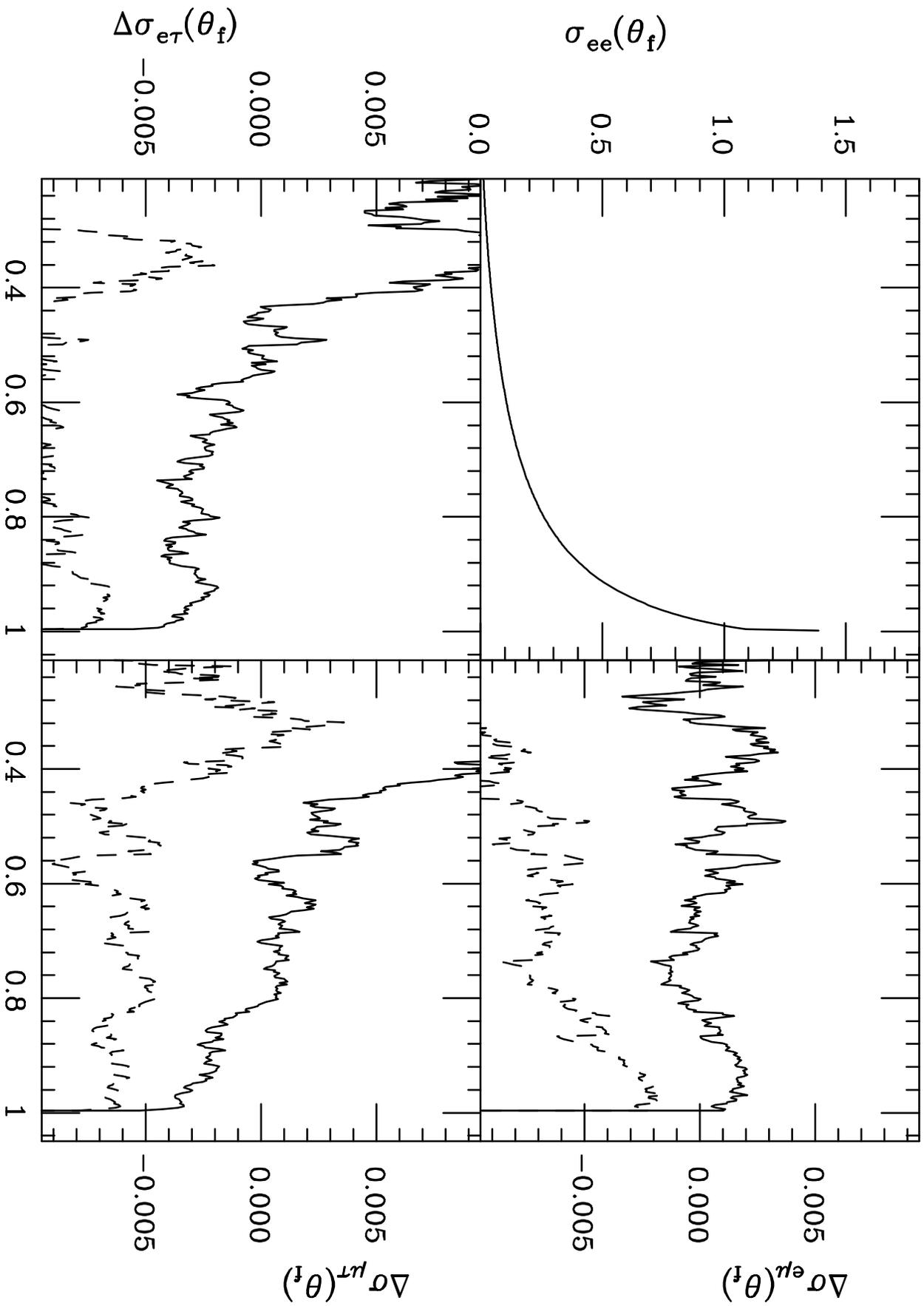
\item [Fig.~2] Cross section for $CC_{jj}$ processes as a function
of the angular cut $\theta_f$. The processes are listed in table \ref{xxcc}
and the definition of the angle $\theta_f$ is given in the caption
of the same table. $\sigma_{jj}$ is the cross section for the $CC_{jj}$
process and $\Delta\sigma_{jk}=(\sigma_{kk}-\sigma_{jj})/\sigma_{jj}$.
Continuos and dashed line refer to the value of $\Delta\sigma_{jk}$ plus 
and minus one standard
deviation respectively.

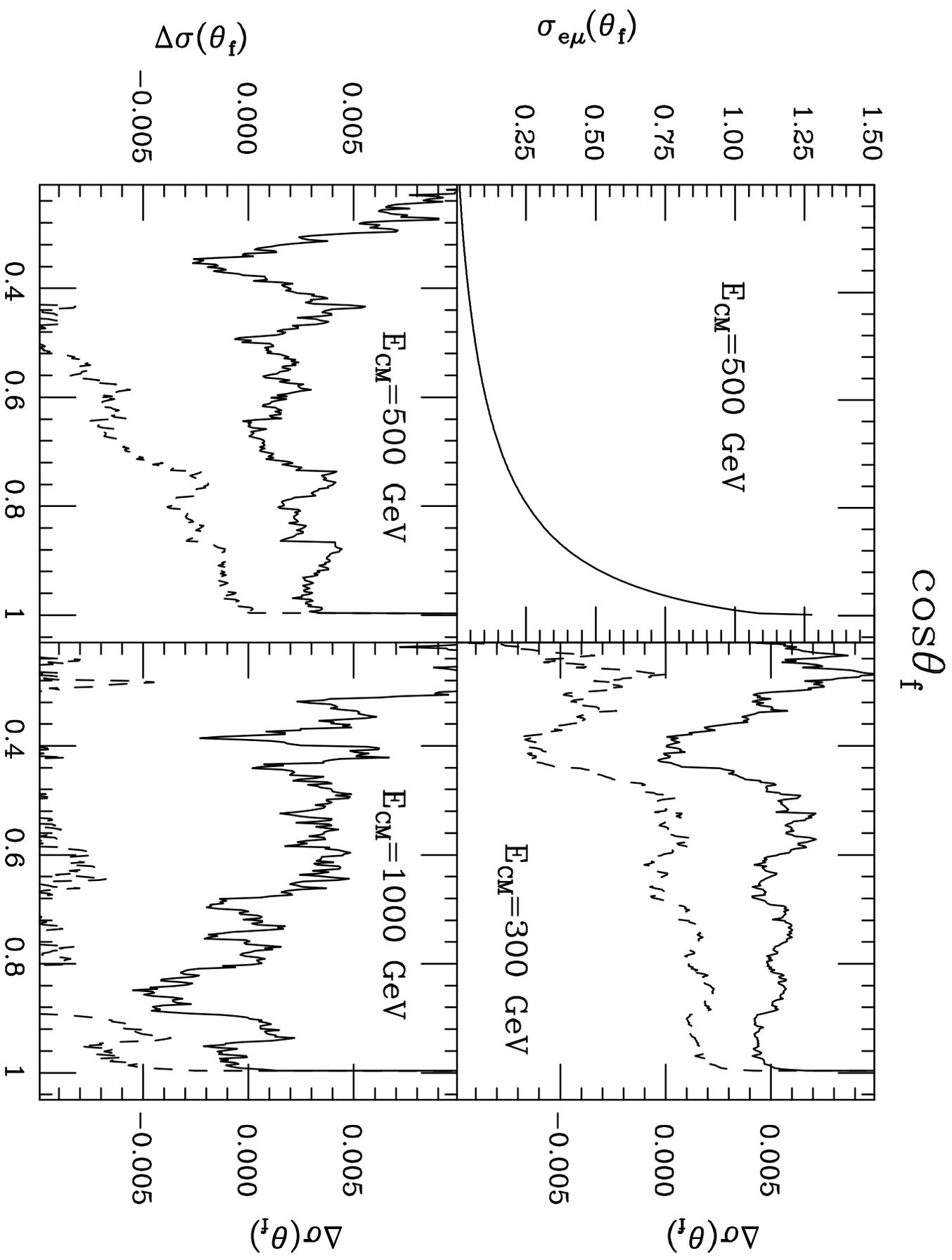
\item [Fig.~3] Cross section for the $CC_{e\mu}$ processes as a function
of the angular cut $\theta_f$. The processes are listed in table \ref{xxcc}
and the definition of the angle $\theta_f$ is given in the caption
of the same table. $\sigma_{e\mu}$ is the cross section for the $CC_{jj}$
process and $\Delta\sigma=(\sigma_{ee}-\sigma_{e\mu})/\sigma_{ee}$.
$E_{CM}$ is the total energy in the center of mass.
Continuos and dashed line refer to the value of $\Delta\sigma$ plus 
and minus one standard
deviation respectively.

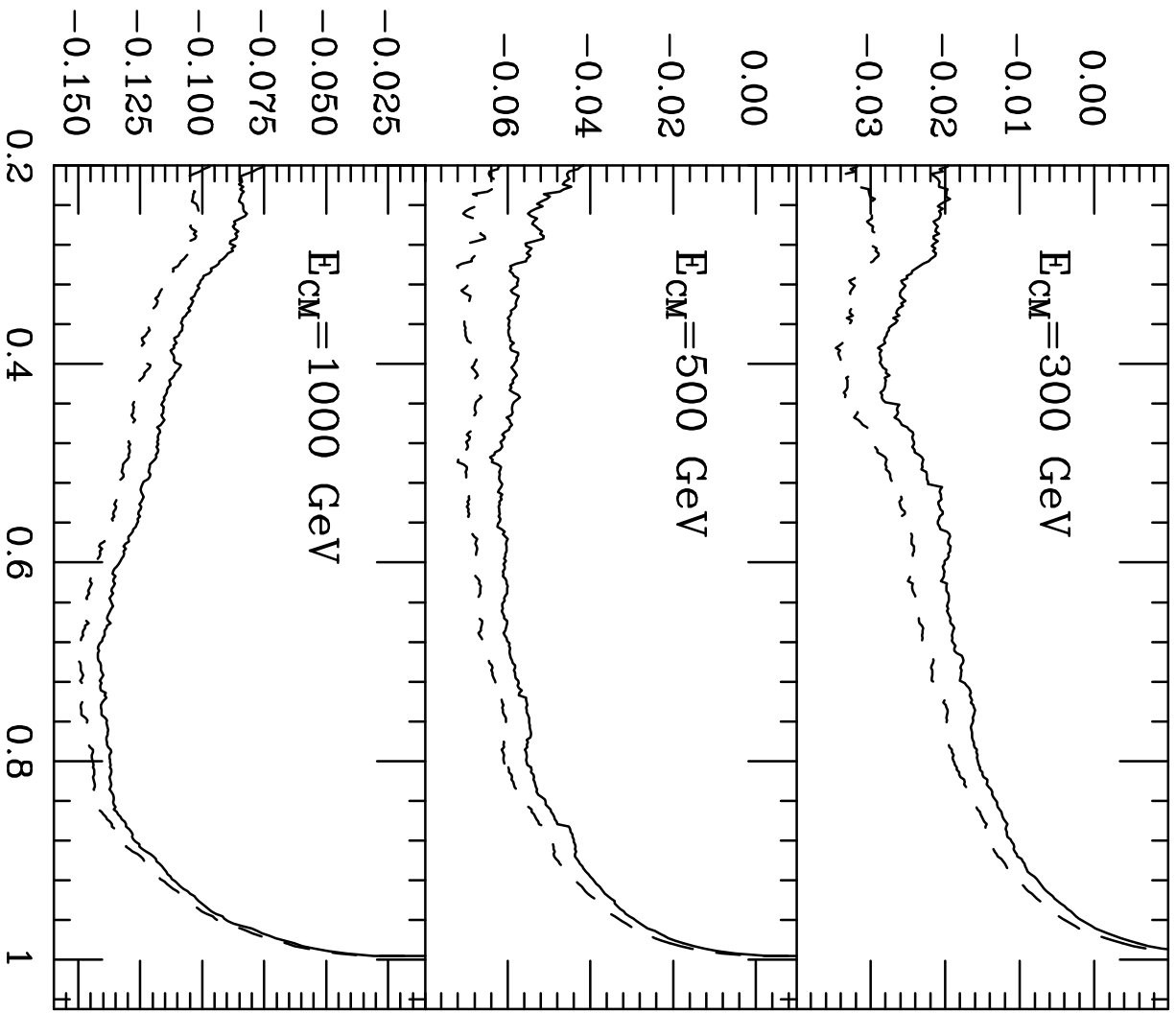
\item [Fig.~4] Relative difference among 
the full calculation and the resonant approximation
for the cross section
of the $CC_{ee}$ processes 
as a function
of the angular cut $\theta_f$. The processes are listed in table \ref{xxcc}
and the definition of the angle $\theta_f$ is given in the caption
of the same table. 
$\Delta\sigma=(\sigma_{ee}-\sigma_{resonant})/\sigma_{ee}$.
$E_{CM}$ is the total energy in the center of mass.
The resonant approximation amounts to consider only the diagrams
$ra$ and $rb$ in fig.~\feyn.
Continuos and dashed line refer to the value of $\Delta\sigma$ plus 
and minus one standard
deviation respectively.

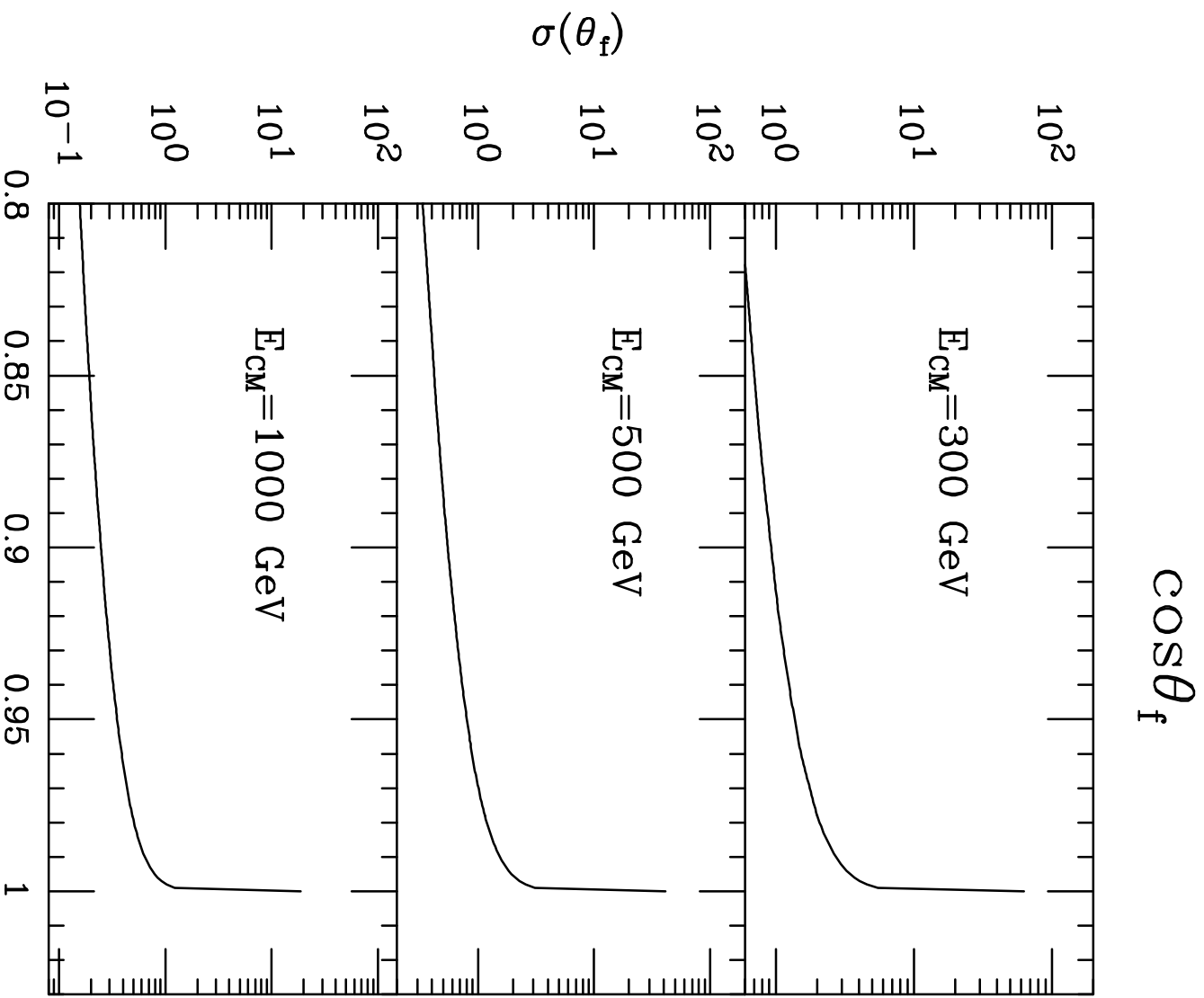
\item [Fig.~5] Cross section in femtobarns for the 
$NC_{e}$ process at a center of mass energies
of 300, 500 and 1000 GeV as a function of the angular cut $\theta_f$.
$E_{CM}$ is the center of mass energy and
the definition of $\theta_f$ is given in
the caption of table \ref{xxnca}. 

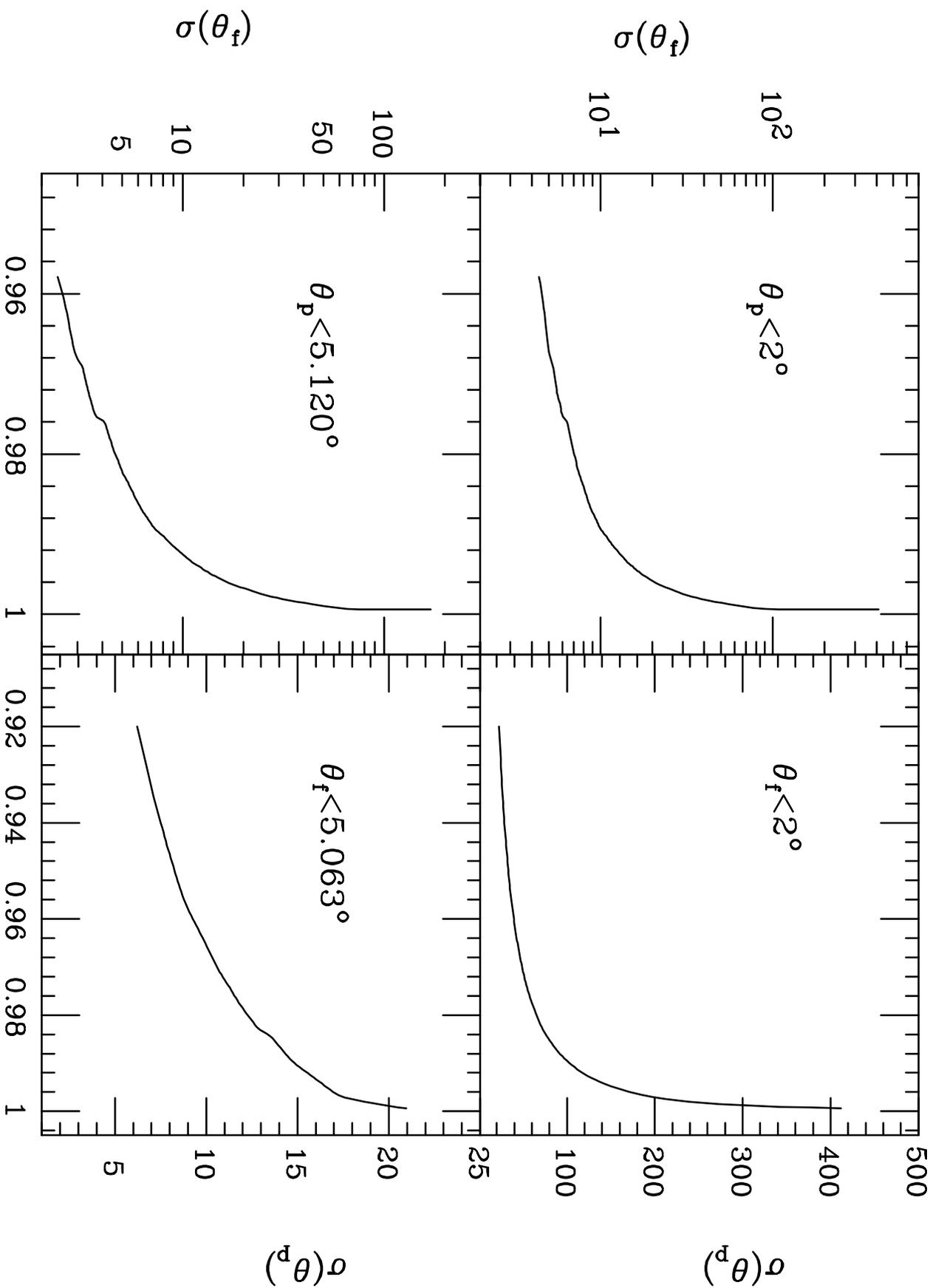
\item [Fig.~6] Cross section in femtobarns for the 
$NC_{e\mu}$ process at a center of mass energy
of 500 GeV as a function of the angular cut $\theta_f$ ($\theta_p$)
at fixed values of the angular cut $\theta_p$ ($\theta_f$).
The definitions of $\theta_p$ and $\theta_f$ are given in
the caption of table \ref{xxncb}.

\end{list}

\end{document}